\documentclass[]{emulateapj}

\def\spose#1{\hbox to 0pt{#1\hss}}
\def\ltsimm{\mathrel{\spose{\lower 3pt\hbox{$\sim$}}
        \raise 2.0pt\hbox{$<$}}}
\def\gtsimm{\mathrel{\spose{\lower 3pt\hbox{$\sim$}}
        \raise 2.0pt\hbox{$>$}}}
\def\Mdot{\hbox{${\dot M}$}}
\def\vinfty{\hbox{${v_{\infty}}$} \,}
\def\km{{\rm\thinspace km}}
\def\cm{{\rm\thinspace cm}}

\def\s{{\rm\thinspace s}}
\def\yr{{\rm\thinspace yr}}
\def\g{{\rm\thinspace g}}
\def\kmps{\hbox{${\rm\km\s^{-1}}$}}

\def\erg{{\rm\thinspace erg}}

\def\Hz{{\rm\thinspace Hz}}

\def\ster{{\rm\thinspace ster}}
\def\ergps{\hbox{${\rm\erg\s^{-1}\,}$}}

\def\Msol{\hbox{${\rm\thinspace M_{\odot}}$}}

\def\Msolpyr{\hbox{${\rm\Msol\yr^{-1}}$}}
\def\pcm{\hbox{${\rm\cm^{-1}\,}$}}
\def\pcm2{\hbox{${\rm\cm^{-2}\,}$}}
\def\pcm3{\hbox{${\rm\cm^{-3}\,}$}}
\def\ergpscm3Hz{\hbox{${\rm\ergps\cm^{-3}\Hz^{-1}\,}$}}
\def\ergpscm3Hzster{\hbox{${\rm\ergps\cm^{-3}\Hz^{-1}\ster^{-1}\,}$}}
\def\gpcm3{\hbox{${\rm\g\cm^{-3}\,}$}}
\def\ergpcm2{\hbox{${\rm\erg\cm^{-2}\,}$}}
\def\ergpcm3{\hbox{${\rm\erg\cm^{-3}\,}$}}
\def\phpscm2{\hbox{${\rm photons\s^{-1}\cm^{-2}\,}$}}

\shorttitle{Clump Destruction in Adiabatic Colliding Winds}

\begin{document}

\title{A Clumping Independent Diagnostic of Stellar Mass-loss Rates: \\ Rapid Clump Destruction in Adiabatic Colliding Winds}

\author{J. M. Pittard\altaffilmark{1}}
\altaffiltext{1}{School of Physics and Astronomy, The University of Leeds,
Leeds, UK: jmp@ast.leeds.ac.uk.}

\begin{abstract}
Clumping in hot star winds can significantly affect estimates of
mass-loss rates, the inferred evolution of the star and the
environmental impact of the wind.  A hydrodynamical simulation of a
colliding winds binary (CWB) with clumpy winds reveals that the clumps
are rapidly destroyed after passing through the confining shocks of
the wind-wind collision region (WCR) for reasonable parameters of the
clumps if the flow in the WCR is adiabatic.  Despite large density and
temperature fluctuations in the post-shock gas, the overall effect of
the interaction is to smooth the existing structure in the
winds. Averaged over the entire interaction region, the resulting
X-ray emission is very similar to that from the collision of smooth
winds. The insensitivity of the X-ray emission to clumping suggests it
is an excellent diagnostic of the stellar mass-loss rates ($\Mdot$) in
wide CWBs, and may prove to be a useful addition to existing
techniques for deriving $\Mdot$, many of which are extremely sensitive
to clumping. Clumpy winds also have implications for a variety of
phenomena at the WCR: particle acceleration may occur {\em throughout}
the WCR due to supersonic MHD turbulence, re-acceleration at multiple
shocks, and re-connection; a statistical description of the properties
of the WCR may be required for studies of non-equilibrium ionization
and the rate of electron heating; and the physical mixing of the two
winds will be enhanced, as seems necessary to trigger dust formation.
\end{abstract}


\section{INTRODUCTION}
There is now considerable observational evidence for a high degree of
structure, or ``clumping'', in hot stellar winds
\citep[e.g.][]{Moffat:1988,Lepine:2000}. Clumping affects diagnostics
of mass-loss rates which are sensitive to the square of the density
(e.g., free-free radio, infra-red continuum, H$\alpha$).  If not taken
into account, the inferred mass-loss rates may be substantially above
their actual values.  For Wolf-Rayet (WR) stars, the assumption of a
smooth wind typically leads to a factor of 3 overestimate. Recent
studies are now indicating that the winds of main sequence stars may
be even more structured, and downward revisions in $\Mdot$ by factors
of $3-10$ or more have been suggested
\citep[e.g.,][]{Bouret:2005,Fullerton:2006,Puls:2006}. Such
substantial reductions have a dramatic effect on the evolution and
environmental impact of massive stars, and may also be needed to
explain the near-symmetry of X-ray lines \citep{Owocki:2006}.

Several methods for determining mass-loss rates which are not sensitive to
$\rho^{2}$  are unique to binary systems. First, if the
period of the binary is short enough, and the mass-loss rates high
enough, $\Mdot$ can be determined from the observed change in the
orbital period.  Though this has been applied to the WR binary
V444~Cyg \citep[e.g.][]{Antokhin:1995}, it has limited applicability.
Observed changes in polarization and atmospheric continuum eclipses
can also be used to determine mass-loss rates 
\citep{StLouis:1988,Lamontagne:1996}.

Measurements of $\Mdot$ have also been made by comparing the observed
X-ray flux arising from the WCR to predictions from hydrodynamical
models of this interaction
\citep{Stevens:1996,Pittard:2002,Pittard:2006b}. In all of these
models homogeneous winds were assumed, but this work shows that structured
winds are rapidly smoothed out in the WCR of {\em wide} binaries: hence, the
X-ray emission, despite being sensitive to $\rho^{2}$, is an excellent
diagnostic of the actual mass-loss rates. Further implications of wind
structure on the interaction region are discussed in
\S~\ref{sec:implications}.

\section{CLUMP-WCR INTERACTION}
\label{sec:clump_properties}
The survival time of a clump within the WCR depends predominantly on
its size and density contrast to the mean flow. The low amplitude of
detected variability argues for a large number of clumps, each of
which is likely to have a quite small spatial scale \citep[e.g.][and
references therein]{Eversberg:1998,Marchenko:2006}. If the interclump
medium is devoid of material, the density contrast of the clumps
relative to the corresponding smooth flow is inversely related to the
volume filling factor, $f_{\rm v}$, which the clumps occupy. $f_{\rm
v}$ initially decreases with radius (i.e. the wind becomes
increasingly clumpy), reaches a minimum at around $10-15 \;R_{*}$, and
then declines as the wind slowly smooths out \citep{Puls:2006}. At
very large distances, \citet{Runacres:2005} find that $f_{\rm v} \sim
0.25$.

Clumps which pass into the WCR lose mass primarily through dynamical
instabilities (e.g. Kelvin-Helmholtz). The destruction timescale of
non-radiative clumps can be parameterized as $t_{\rm d} = \epsilon
r_{\rm c}/v_{\rm s}$ \citep{Klein:1994}, where $r_{\rm c}$
is the clump radius and $v_{\rm s}$ is the shock velocity (for the
stationary shocks in CWBs, $v_{\rm s}$ is equal to the normal
component of the pre-shock wind speed).  $\epsilon \approx 3.5$ for a
density ratio between the clump and inter-clump wind of order
$10-100$. A detailed review of clump destruction processes, including
the effects of radiative cooling and magnetic fields, can be found in
\citet{Pittard:2006c}.

Consider a clump moving along the line of centers between the stars.
As the clump passes through one of the shocks confining the WCR it is
decelerated less than the inter-clump material, and pushes the
confining shock into the WCR. A lower limit on the timescale for the
clump to half cross the WCR is $t_{\rm cd} = \Delta r/v_{\rm s}$,
where $\Delta r$ is half the distance between the confining
shocks. Clumps will be destroyed before they reach this half-way point
when $t_{\rm d}/t_{\rm cd} = \epsilon r_{\rm c}/\Delta r < 1$.
Since it is expected that $r_{\rm c} \ltsimm \Delta r$, clumps should
be rapidly destroyed within the WCR.

\section{HYDRODYNAMICAL SIMULATIONS}
The long-period CWB \object[HD 193793]{WR\thinspace140} forms the
basis of this investigation into the effects of clumpy winds on the
interaction region.  WR\thinspace140 is the achetype of long-period CWBs,
exhibiting dramatic variations in its X-ray \citep{Pollock:2005} and
radio emission \citep[][]{Dougherty:2005} modulated by its highly
eccentric orbit. The emission from radio to TeV energies has recently
been modelled by \citet{Pittard:2006b}.

There have been two previous studies of the clumpy colliding winds in
WR\thinspace140. \citet{Walder:2002} examined a simulation with the
stars near periastron. WR material both in and between the clumps
rapidly cools, and is compressed to high
densities. \citet{Aleksandrova:2000} considered the transition between
periastron and apastron in an attempt to explain the variation of the
X-ray flux with the stellar separation.

In this work the effect of clumpy winds on the WCR is examined when
the stars are at apastron (i.e. the stellar separation $D=4.7 \times
10^{14}\;{\rm cm}$). The mass-loss rates and terminal wind speeds
adopted are: $\Mdot_{\rm WR} = 4.3 \times 10^{-5}\;\Msolpyr$,
$\Mdot_{\rm O} = 8.0 \times 10^{-7}\;\Msolpyr$, $\vinfty_{\rm WR} =
2860 \;\kmps$, $\vinfty_{\rm O} = 3100 \;\kmps$. In order to focus
attention on the apex of the WCR, which is bent sharply around the O
star, the WR star is positioned off the 2D hydrodynamic grid, which is
axisymmetric and contains $1650\times660$ cells. The clumps are added
to the flow in annuli around each star at specific time intervals. For
simplicity the clumps are assigned the wind terminal speed, and have
radii proportional to their distance from their star.  The clump
radius, $r_{\rm c}$, is $0.005\;r$ in the WR wind, and $0.02\;r$ in
the O wind, so that the clumps have $\gtsimm 10$ cells across their
radius when interacting with the WCR.  The clumps are given a density
contrast of 10 with respect to the interclump medium, and the clump
and interclump medium are assumed to contain equal mass (hence $f_{\rm
v} = 1/11$). The inter-clump medium is perfectly smooth. A simulation
with smooth winds reveals that the width of the WCR on the line
between the centers of the stars is $0.0695\;D$. Hence $t_{\rm
d}/t_{\rm cd} \approx 0.5$ (0.25) for the clumps in the WR (O) wind,
and the WCR should quickly smooth out these inhomogeneities.

The hydrodynamical code is second-order accurate in space and time
\citep{Falle:1996}. Optically thin radiative cooling is included,
though radiation losses are negligible. Heat conduction is not
explicitly included, and is likely to be strongly inhibited by the
magnetic field revealed by the synchrotron emission. The clumps are
initially spherical, but the velocity dispersion in the radial and
transverse directions may differ. $v_{\theta}/v_{\rm r} \approx 4$ 
for the clumps in the O wind, so that they naturally
``pancake'' as they are advected by the wind. Recent models of X-ray
wind absorption support radially compressed rather than spherical
clumps \citep{Oskinova:2006}. The density contrast of the clumps is
also reduced by numerical diffusion. These two effects reduce
the survival time of the clumps, but are not thought to be significant.

\begin{figure}
\begin{center}
\includegraphics[angle=-90,width=76mm]{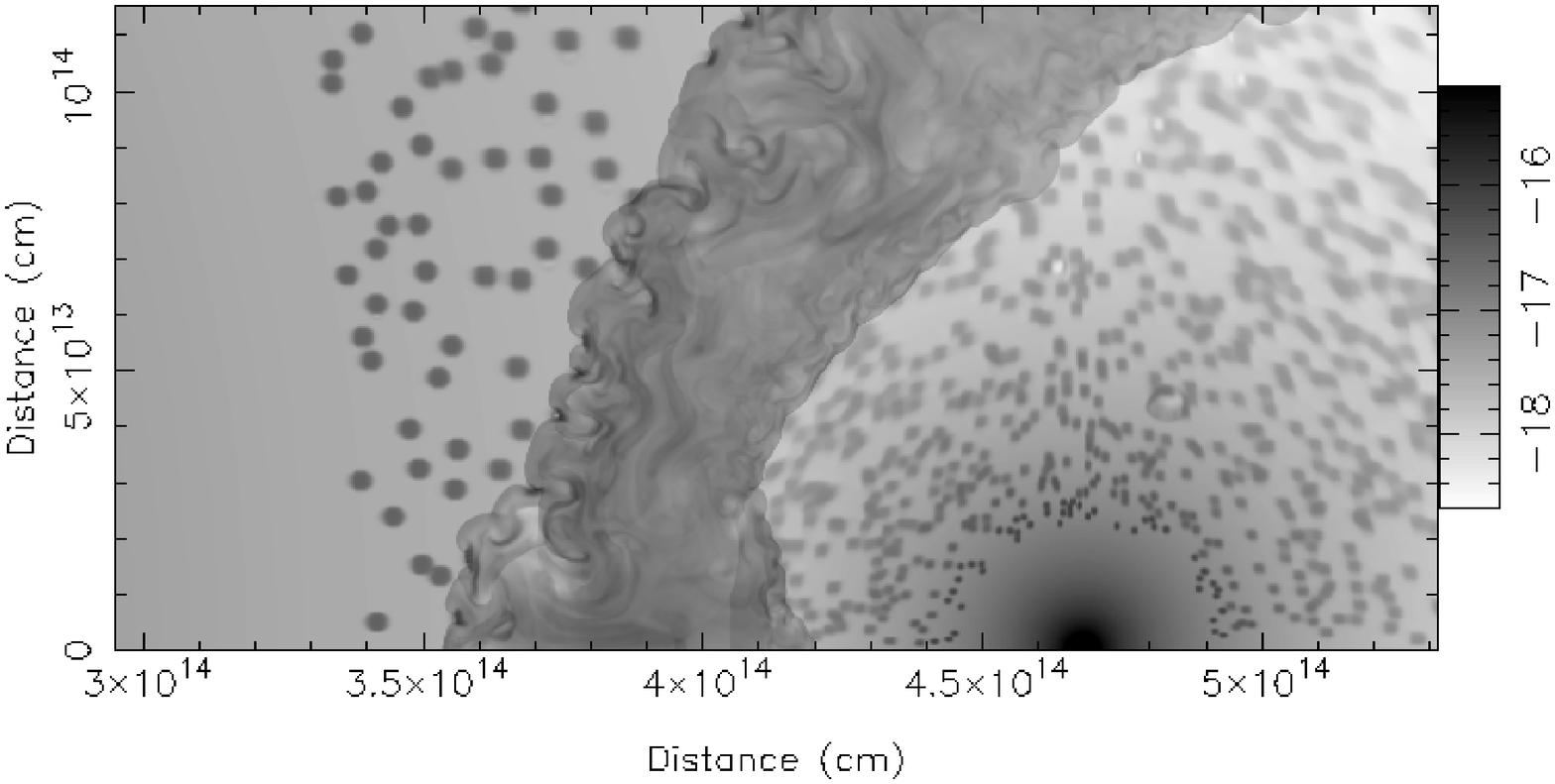}\\
\includegraphics[angle=-90,width=76mm]{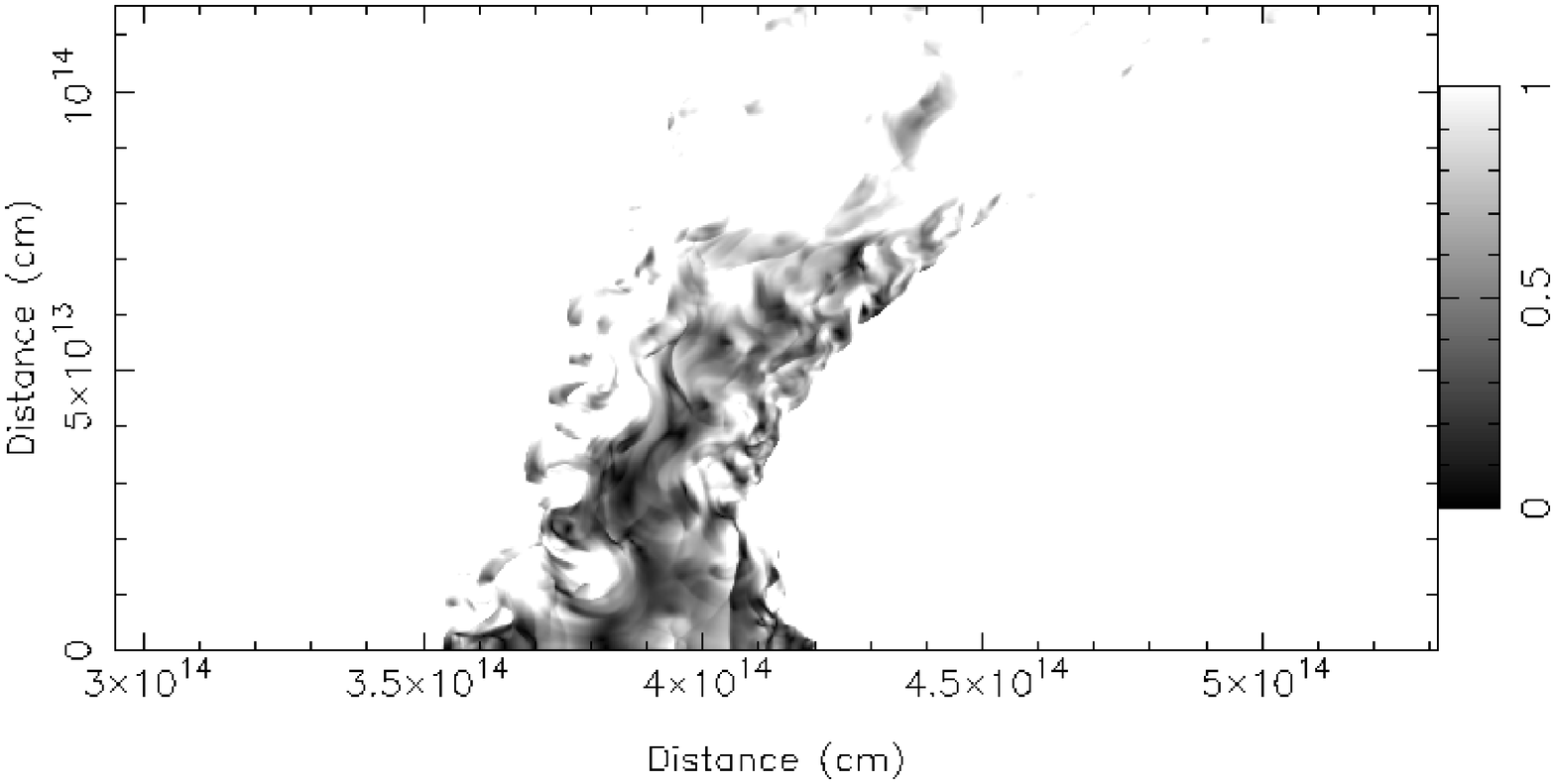}
%wr140_clumpy_run21_t19.0_ptvn_gray_mach.ps}
\caption{a) Logarithmic density plot (${\rm log_{10}\; g\;cm^{-3}}$)
from an axisymmetric simulation of colliding clumped winds in
WR\thinspace140. b) As a) but showing the Mach number of the gas. 
Supersonic gas is white. Due to the pressure gradients within the WCR, 
the gas, on average, becomes supersonic as it leaves the system.}
\label{fig:wcr}
\end{center}
\end{figure}

\begin{figure}
\begin{center}
\plotone{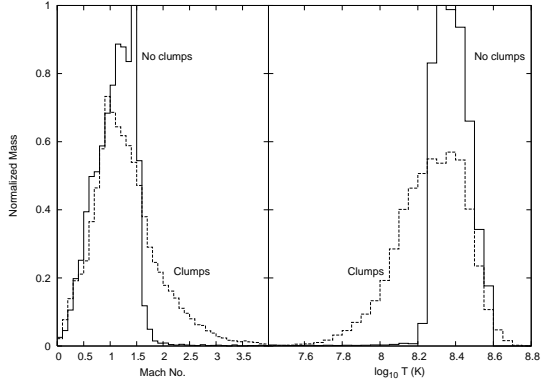}
\caption{a) Comparison of mass as a function of Mach number from
simulations with homogeneous (solid line) and clumpy (dashed
line) winds. b) As a) but as a function of temperature.}
\label{fig:massvs}
\end{center}
\end{figure}

A density plot of the WCR is shown in Fig.~\ref{fig:wcr}a). It is
immediately evident that the process of clump destruction induces a
multitude of large and small scale motions within the WCR. This
``turbulence'', some of which is supersonic (Figs.~\ref{fig:wcr}b
and~\ref{fig:massvs}a), puffs-up the WCR, so that its volume is 25\%
larger than the homogeneous case.  Temperatures higher than in
the smooth-wind case occur when the WCR expands upstream into the
relatively low ram-pressure of the interclump medium, and also behind
the bow-shocks driven ahead of the clumps. In contrast, the interiors
of clumps are initially heated to significantly lower temperatures
as shocks driven into them are much slower (shocks driven into 
clumps in the WR wind near the line of centers have speeds $\sim
1000\;\kmps$, and produce temperatures of $\sim 4 \times 10^{7}\;{\rm
K}$). Thus, there is a wider distribution of temperatures within the
WCR than in the smooth winds case (Fig.~\ref{fig:massvs}b).

As cool material within recently-shocked clumps moves deeper into the
WCR it is heated by secondary shocks and through mixing with hotter
plasma. At any particular instant, the majority of the mass within the
WCR is heated to temperatures similar to those that exist in the
smooth winds simulation: the mass-weighted mean temperature of gas
within the WCR (and on the grid) is $2.4 \times 10^{8}
\;{\rm K}$ and $2.0 \times 10^{8}\;{\rm K}$ for the smooth and
structured winds cases respectively. The average density in the WCR is
similar to the smooth winds case, and is not markedly different with
90\% of the mass in clumps.

Although the WCR is clearly not smooth, the wind-wind collision
decreases the small-scale structure in the winds.  Clump disruption is
mostly through the development of large scale perturbations, while the
mixing of the clump and interclump material is affected by smaller
scale motions.  The global nature of the simulation inevitably leads
to relatively poor spatial resolution at the scale of individual
clumps, and is likely to enhance the mixing due to non-negligible
numerical diffusion.  In contrast, the lack of a third dimension in
these simulations (the imposition of axisymmetry means that the clumps
are donut shaped) should slow the rate of clump destruction as there
is one less degree of freedom for dynamical instabilities.  Resolution
tests and comparison with previous work on shock-cloud interactions
indicate that these effects do not have a serious impact on the
results.

In reality, clumps will possess a variety of density contrasts and
sizes - large, dense clumps will survive longer as distinct entities
within the WCR than smaller, less dense clumps, though denser
clumps may on average be smaller and vice-versa \citep{Moffat:1994}.
If $t_{\rm d}/t_{\rm cd} >> 1$, clumps could, in theory, pass
completely through the WCR and into the pre-shock wind of the
companion star, but this is unlikely to occur in the wide, adiabatic
systems considered here.

\subsection{Determining Mass-loss Rates} 
\begin{figure}
\begin{center}
\plotone{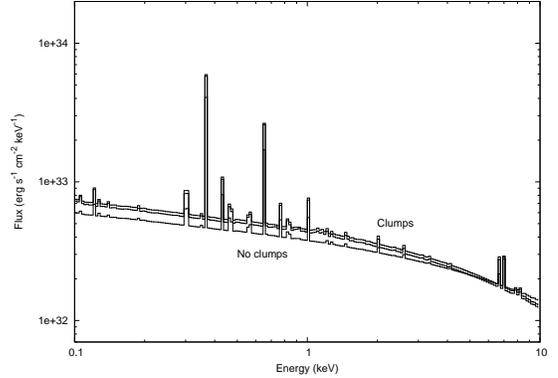}
\caption{Comparison of the X-ray emission from simulations with
homogeneous (thick line) and clumpy (thin lines) winds. The latter is
slightly softer, and shows only weak variability ($\ltsimm 5$\%),
despite the fact that the hydrodynamic grid is not large enough to
capture all of the emission (note the lack of lines below 1~keV). The
emission was calculated using the MEKAL code \citep[][and references
therein]{Mewe:1995}, for optically thin thermal plasma, ionization
equilibrium, and identical ion and electron temperatues.}
\label{fig:xray}
\end{center}
\end{figure}

Since the wind structure is rapidly smoothed, it is not surprising
that the X-ray emission is similar to the homogeneous case
(Fig.~\ref{fig:xray}).  The mean continuum flux at 1~keV is only 17\%
higher in the clumpy winds simulation, which translates into an 8\%
overestimate of the mass-loss rates ($f_{\rm x} \propto
\Mdot^{2}$). Clearly, there is potential to use the X-ray emission
from the WCR as a diagnostic of the stellar mass-loss rates.

However, the structure within the WCR and the resulting X-ray emission
depends on the properties of the clumps. Further simulations
(Pittard, in preparation) reveal that the emission measure ($\int
n^{2} dV$) appears to increase with the clump density contrast: an
identical simulation with a density contrast of 100 yields a
56\% overestimate of $\Mdot$ for each star, but the density contrast
is unlikely to be this high in wide CWB systems. Instead, the major
sources of uncertainty are likely to be the abundances and the wind
momentum ratio, $\eta$ \citep[as the emission is nearly degenerate
between $\eta$ and the $\Mdot$'s - see][]{Pittard:2006b}.
Nevertheless, X-ray derived mass-loss rates should be accurate to within
a factor of 2. Substantially higher precision may be obtained if
$\eta$ and the wind abundances are strongly constrained.

The level of X-ray absorption by the intervening wind(s) can also, in
principle, be used to determine mass-loss rates. For the wide binaries
considered here the clumps are likely to be optically thin, so that
the optical depth will be identical to the smooth winds case. However,
again there is a degeneracy with $\eta$ if the line-of-sight into the
system is not well constrained, in which case it may be best to simply
match the X-ray flux at energies above those susceptible to absorption
\citep{Pittard:2006b}.

\section{FURTHER IMPLICATIONS}
\label{sec:implications}
The interaction of clumpy, as opposed to homogeneous, winds has 
implications for a variety of phenomena which occur at or within the WCR.
 
\subsection{Particle Acceleration and Synchrotron Emission}
The radio synchrotron emission from wide CWBs has recently been
modelled assuming that relativistic electrons are created by diffusive
shock acceleration (DSA) at the shocks confining the WCR
\citep{Dougherty:2003,Pittard:2006a,Pittard:2006b}. In the case of
WR\thinspace140, \citet{Pittard:2006b} find that the non-thermal (NT)
electron energy distribution is harder than the canonical DSA value,
i.e. $p < 2$. While there are many possible explanations, the results
presented above hint at several mechanisms. First, NT
particles may be accelerated via the second-order Fermi process
resulting from clump-induced MHD turbulence within the WCR
\citep[e.g.][]{Scott:1975}.  Second, particles accelerated at the
confining shocks may be re-accelerated at multiple (weak) shocks
within the WCR \citep{Schneider:1993}. A third possibility is magnetic
reconnection, which probably occurs throughout the volume of the
turbulent WCR, and not just at a hypothetical contact
discontinuity. Reconnection may also provide additional energy for
generating and maintaining the magnetic fluctuations which drive
stochastic acceleration. If any of these mechanisms is dominant,
models which impose particle acceleration only at the confining shocks
will need to be revised accordingly. In addition, the magnetic field
within the WCR is likely to be highly ``tangled'' due to the turbulent
motions of the gas inside. This was a central assumption in the recent
models mentioned above, and means that the asymmetric radio lightcurve
of WR\thinspace140 cannot be explained by the angular dependence of
the synchrotron emission process.

\subsection{Plasma Timescales}
There are a variety of timescales within the WCR which may be modified
if the stellar winds are clumpy \citep[e.g.,][]{Walder:2002}. In wide
CWB systems, the shocks are collisionless, and the electron and ion
temperatures may differ \citep[an effect which appears to be sensitive
to the shock speed, e.g.][]{Hwang:2002}. Equilibration subsequently
occurs through Coulomb collisions, proceeding faster where the density
is higher, such as in material within the shocked clumps.  The
spectral hardness is sensitive to this process, but the overall flux
is not dramatically changed, and it is unlikely to have a significant
impact on mass-loss rate determinations.

Ionization equilibrium may also take a significant time to occur.
Direct evidence for non-equilibrium ionization is seen in
WR\thinspace140 \citep{Pollock:2005}, and models which assume
ionization equilibrium fail to reproduce observed X-ray line profiles
\citep{Henley:2005}.  Material originally within clumps will ionize
quicker than interclump material, but to lower stages due to the
reduced post-shock temperatures. The relevant timescale to obtain the
highest ionization stages then becomes the speed at which further
heating occurs in the downstream flow, through compressions, shocks,
and mixing. The details are again sensitive to the clump properties,
but the continuum emission, and thus estimates of $\Mdot$, should not
be strong affected.

\subsection{Dust Formation}
The highly turbulent interior of the WCR shown in Fig.~\ref{fig:wcr}
enhances the mixing between the two winds. Such mixing may be necessary in 
order that carbon-rich WR material and hydrogen-rich O-star material
can form dust within the WCR 
\citep[e.g.][]{Walder:2002}.

\section{SUMMARY}
The interaction of clumpy stellar winds in massive binary systems
creates a highly turbulent wind-wind collision region in the adiabatic
limit. The lifetime of clumps within the WCR depends on their density
contrast and size. Clumps with a density contrast of 10 and radii a
few times smaller than the half-width of the WCR on the line of centers
between the stars are rapidly destroyed. Material originally within
the clumps is then vigourously mixed into the surrounding flow.

The stochastic impact of clumps on the WCR and their subsequent
destruction creates significant density and temperature fluctuations
within the WCR, but the global X-ray emission can be remarkably
similar to the smooth winds case. The X-ray emission is then an
effective, clumping-independent, measure of the stellar mass-loss
rates, especially if the wind momentum ratio is known. The small
number of CWB systems suitable for such an analysis is countered by
the potential accuracy that can be obtained. Each time this method has
been used to determine mass-loss rates, values lower than those in the
literature were inferred. Although only a small part of parameter
space is explored here, any reasonable parameters for the clumps in
the wide, adiabatic, systems considered here should lead to their
rapid dissolution.
 
Turbulence and weak shocks within the WCR provide mechanisms for
obtaining hard NT particle spectra. The timescale to obtain
high ionization stages may be controlled by the rapidity of clump
destruction, and enhanced mixing of the winds will aid dust
formation.

\acknowledgments
The author thanks the Royal Society for funding, Sam Falle
for use of his hydro code, and Tom Hartquist, Sean Dougherty, 
Ian Stevens and the referee for comments.

\end{document}